\documentstyle[12pt,epsfig]{article}
\def\lbldef#1#2{\expandafter\gdef\csname #1\endcsname {#2}}

\def\href#1#2{#2}  
\begin{document}
\baselineskip=15.5pt
\pagestyle{plain}
\setcounter{page}{1}
%\renewcommand{\thefootnote}{\fnsymbol{footnote}}
%--------+---------+---------+---------+---------+---------+---------+
%Title page

\begin{titlepage}

\begin{flushright}
CERN-TH/99-329\\
hep-th/9910244
\end{flushright}
\vspace{10 mm}

\begin{center}
{\Large  Black Hole Thermodynamics and Two-Dimensional Dilaton 
Gravity Theory}

\vspace{5mm}

\end{center}

\vspace{5 mm}

\begin{center}
{\large Donam Youm\footnote{Donam.Youm@cern.ch}}

\vspace{3mm}

Theory Division, CERN, CH-1211, Geneva 23, Switzerland

\end{center}

\vspace{1cm}

\begin{center}
{\large Abstract}
\end{center}

\noindent

We relate various black hole solutions in the near-horizon region to 
black hole solutions in two-dimensional dilaton gravity theories in 
order to argue that thermodynamics of black holes in $D\geq 4$ can be 
effectively described by  thermodynamics of black holes in 
two-dimensional dilaton gravity theories.  We show that the Bekenstein-Hawking 
entropies of single-charged dilatonic black holes and dilatonic 
$p$-branes with an arbitrary dilaton coupling parameter in arbitrary 
spacetime dimensions are exactly reproduced by the Bekenstein-Hawking entropy 
of the two-dimensional black hole in the associated two-dimensional dilaton 
gravity model.  We comment that thermodynamics of non-extreme stringy 
four-dimensional black hole with four charges and five-dimensional black 
hole with three charges may be effectively described by thermodynamics of 
the black hole solutions with constant dilaton field in two-dimensional 
dilaton gravity theories.

\vspace{1cm}
\begin{flushleft}
CERN-TH/99-329\\
October, 1999
\end{flushleft}
\end{titlepage}
\newpage

\section{Introduction}

The recent development in string duality and the observation \cite{pol} that 
D-branes can carry charges of form potentials in the RR sector of string 
theories enabled us to address fundamental issues in quantum gravity such as 
statistical interpretation of black hole entropy within the framework of 
string theory.  By applying the D-brane counting technique, pioneered in 
Ref. \cite{sv}, it has been successful in reproducing the 
Bekenstein-Hawking entropies of black hole solutions in string theories.  
Generally, the D-brane counting technique can be applied to the BPS and 
the near-BPS cases and few other special cases such as non-BPS extreme 
rotating black hole in four dimensions. 

Holographic principle \cite{tho,sus1,sus2} also has contributed to the 
microscopic interpretation of the Bekenstein-Hawking entropy.  Brown and 
Henneaux \cite{bh} showed that the asymptotic symmetry group of 
AdS$_3$ space is the conformal group in two dimensional spacetime at the 
boundary.  By using this idea, Carlip \cite{car1,car2} and Strominger 
\cite{str} were able to exactly reproduce the Bekenstein-Hawking entropy 
of the three-dimensional constant curvature black hole of Ban\~ ados, 
Teitelboim and Zanelli (BTZ) \cite{btz} by counting microscopic degrees 
of freedom of the conformal field theory at the boundary.  An important 
observation \cite{hyu} by Hyun that (uplifted) black hole solutions in 
string theory can be put in the form of the product of the BTZ black hole 
and a sphere through a series of $U$-duality transformations enabled the 
microscopic counting of non-extreme stringy black holes in four and five 
dimensions by applying the result of Carlip and Strominger (for example,  
Ref. \cite{ss}).  

As pointed out in Ref. \cite{hyu}, higher-dimensional black hole 
solutions in string theories can also be related to two-dimensional black 
hole solutions through $U$-duality transformations.  The four- and 
the five-dimensional non-extreme black holes in string theory are 
related \cite{teo,cds1,cds2,klp,cdn} to two-dimensional black holes in  
the Jackiw-Teitelboim (JT) model \cite{jac,tei} and the 
Callan-Giddings-Harvey-Strominger (CGHS) model \cite{cghs} and to the 
two-dimensional charged black hole of McGuigan, Nappi and Yost 
\cite{mny}.  Recently, progress has been made \cite{cdm1,cdm2,nav} in 
reproducing the Bekenstein-Hawking entropies of the two-dimensional black 
holes by using the Cardy's formula \cite{card} for the boundary conformal 
theory.  Furthermore, it is argued in Refs. \cite{carl,sol} that 
the Bekenstein-Hawking entropy of the generic black holes in arbitrary 
dimensional pure gravity theories can be reproduced by the microscopic 
calculation based on the conformal theory associated with the 
two-dimensional subset of the spacetime.  So, it seems that black holes 
in two-dimensional gravity theories, which were originally studied as 
unrealistic toy models for quantum theory of gravity, have closer connection 
with realistic black holes in $D\geq 4$ than it was originally thought.

It is the purpose of this paper to relate various black hole 
solutions in $D\geq 4$ to black hole solutions in two-dimensional 
dilaton gravity theories.  In section 2, we relate a single-charged 
dilatonic black hole solution in $D\geq 4$ with an arbitrary dilaton 
coupling parameter to a black hole solution in a two-dimensional dilaton 
gravity theory.  We find that the Bekenstein-Hawking entropies of these 
two black holes are the same in the large charge limit or the near-extremal 
limit.  In section 3, we argue that the thermodynamics of a dilatonic 
$p$-brane with an arbitrary dilaton couple parameter in arbitrary spacetime 
dimensions can also be described by a black hole solution in a 
two-dimensional dilaton gravity theory.  In section 4, we comment that 
thermodynamics of the non-extreme four-dimensional stringy black hole 
with four charges and the non-extreme five-dimensional stringy black hole 
with three charges may be more naturally described by the black holes 
with constant dilaton field in two-dimensional dilaton gravity theories.

\section{Single-Charged Dilatonic Black Holes}

In this section, we consider single-charged dilatonic black hole solutions 
with an arbitrary dilaton coupling parameter $a$ in arbitrary spacetime 
dimensions $D$.  The result in this section can be applied not only 
to all the single-charged dilatonic black holes but also to 
multi-charged black holes in string theories, because in the 
near-horizon limit the actions for the multi-charged black hole 
solutions in string theories can be reduced (with the guidance of the 
explicit solutions in the near-horizon limit) to the actions for the 
two-dimensional dilaton gravity with possibly different forms of the 
dilaton potential term from the one obtained in this section.  The exceptional 
case is the multi-charged stringy black hole solutions with regular BPS 
limit, i.e., the four-dimensional black hole with four charges and the 
five-dimensional black hole with three charges.  In this special 
case, all the scalars of the solution become constant in the 
near-horizon limit and therefore one cannot relate such black hole 
solutions to black hole solutions  with non-trivial dilaton field in 
two-dimensional dilaton gravity models
\footnote{On the other hand, the near-horizon region solutions for 
this case can be related to the BTZ black hole solution when the solutions 
are uplifted to one higher spacetime dimensions, after the $U$-duality 
transformations when necessary.  (Note, however, that in such embeddings of 
black hole solutions the harmonic function for the charge associated with 
the gravitational wave is allowed to not take the near-horizon form and 
therefore some of scalars are not constant.)  The action for the BTZ black 
hole solution can be reduced to a two-dimensional dilaton gravity model action 
upon dimensional reduction.  But in this paper, we shall ignore the 
higher-dimensional origin of black hole solution, just considering the 
Einstein-frame metric itself.  See the second paragraph of the last section 
for more comments.}.  
This exceptional case will be separately discussed in the last section.  

The corresponding Einstein-frame action is 
\begin{equation}
S_E={1\over{2\kappa^2_D}}\int d^Dx\sqrt{-G^E}\left[{\cal R}_{G^E}
-{4\over{D-2}}(\partial\phi)^2-{1\over 4}e^{2a\phi}F^2_2\right],
\label{ein0dilact}
\end{equation}
where $\kappa_D$ is the $D$-dimensional Einstein gravitational constant 
and $F_2$ is the field strength of the $U(1)$ gauge potential $A^{(1)}=
A_Mdx^M$ ($M=0,1,...,D-1$).  The non-extreme black hole solution to the 
field equations of this action is given by
\begin{eqnarray}
ds^2_E&=&-H^{-{{4(D-3)}\over{(D-2)\Delta}}}fdt^2+H^{4\over{(D-2)\Delta}}
\left[f^{-1}dr^2+r^2d\Omega^2_{D-2}\right],
\cr
e^{\phi}&=&H^{{(D-2)a}\over{2\Delta}},\ \ \ \ \ 
A_t=-H^{-1},
\label{eindil0sol}
\end{eqnarray}
where 
\begin{eqnarray}
H&=&1+{{m\sinh^2\alpha}\over{r^{D-3}}}, \ \ \ \ \ 
f=1-{{m}\over{r^{D-3}}},
\cr 
\Delta&=&{{(D-2)a^2}\over{2}}+{{2(D-3)}\over{D-2}}.
\label{bhsoldefs}
\end{eqnarray}
The Bekenstein-Hawking entropy $S_{BH}$ of the dilatonic black hole solution 
(\ref{eindil0sol}) is determined by the surface area $A_H$ of the event 
horizon (located at $r=r_H=m^{1\over{D-3}}$):
\begin{equation}
S_{BH}={{A_H}\over{4G_D}}={{m^{{D-2}\over{D-3}}V_{S^{D-2}}\cosh^{4\over
\Delta}\alpha}\over{4G_D}},
\label{bhentddim}
\end{equation}
where $V_{S^{D-2}}=2\pi^{{D-2}\over 2}/\Gamma({{D-1}\over 2})$ is the volume 
of $S^{D-2}$ with the unit radius.  

One can also think of the dilatonic black hole in $D$ dimensions as being 
magnetically charged under the $(D-2)$-form field strength 
$F_{D-2}$.  The corresponding Einstein-frame action is
\begin{equation}
\tilde{S}_E={1\over{2\kappa^2_D}}\int d^Dx\sqrt{-G^E}\left[{\cal R}_{G^E}
-{4\over{D-2}}(\partial\phi)^2-{1\over{2\cdot(D-2)!}}e^{-2a\phi}F^2_{D-2}
\right].
\label{dualeinact}
\end{equation}
In terms of this ``dual'' field parametrization, the dilatonic 
black hole solution takes the following form:
\begin{eqnarray}
ds^2_E&=&-H^{-{{4(D-3)}\over{(D-2)\Delta}}}fdt^2+H^{4\over{(D-2)\Delta}}
\left[f^{-1}dr^2+r^2d\Omega^2_{D-2}\right],
\cr
e^{\phi}&=&H^{{(D-2)a}\over{2\Delta}},\ \ \ \ \ 
F_{D-2}=\star(dH\wedge dt).
\label{dualbhsol}
\end{eqnarray}

In the so-called dual-frame \cite{dl,gt,dgt,bst}, the spacetime of the BPS 
dilatonic black hole in the near-horizon region takes the AdS$_2\times 
S^{D-2}$ form.  The following dual-frame action is related to 
(\ref{dualeinact}) through the Weyl rescaling transformation $G^E_{MN}=
e^{-{{2a}\over{D-3}}\phi}G^d_{MN}$:
\begin{equation}
\tilde{S}_d={1\over{2\kappa^2_D}}\int d^Dx\sqrt{-G^d}e^{\delta\phi}
\left[{\cal R}_{G^d}+\gamma(\partial\phi)^2-{1\over{2\cdot(D-2)!}}
F^2_{D-2}\right],
\label{dualfract}
\end{equation}
where the parameters in the action are defined as
\begin{equation}
\delta\equiv-{{D-2}\over{D-3}}a,\ \ \ \ \ \ 
\gamma\equiv{{D-1}\over{D-2}}\delta^2-{4\over{D-2}}.
\label{paradefs}
\end{equation}
In the dual-frame, the dilatonic black hole solution takes the 
following form:
\begin{equation}
dx^2_d=-H^{{2\Delta-4(D-3)}\over{(D-3)\Delta}}fdt^2+H^{2\over{D-3}}
\left[f^{-1}dr^2+r^2d\Omega^2_{D-2}\right],
\label{dualfrsol}
\end{equation}
where the dilaton and the $(D-2)$-form field strength take the same 
forms as in Eq. (\ref{dualbhsol}).  In the near-horizon region, the metric 
(\ref{dualfrsol}) is approximated to
\begin{equation}
ds^2_d\approx-\left({{\hat{\mu}}\over{r}}\right)^{2-{{4(D-3)}\over{\Delta}}}
\left(1-{{m}\over{r^{D-3}}}\right)dt^2+\left({{\hat{\mu}}\over{r}}\right)^2
\left(1-{{m}\over{r^{D-3}}}\right)^{-1}dr^2+\hat{\mu}^2d\Omega^2_{D-2},
\label{neardualmet}
\end{equation}
where $\hat{\mu}\equiv (m\sinh^2\alpha)^{1/(D-3)}$, and the dilaton 
and the non-zero component of the $U(1)$ gauge field are approximated to
\begin{equation}
e^{\phi}\approx\left({\hat{\mu}\over r}\right)^{{(D-2)(D-3)a}\over
{2\Delta}},\ \ \ \ \ \ \ \ 
A_t\approx\left({r\over\hat{\mu}}\right)^{D-3}.
\label{dilu1nearhor}
\end{equation}

By compactifying the dual-frame action (\ref{dualfract}) on $S^{D-2}$ 
with the radius $\hat{\mu}$, one obtains the following two-dimensional 
effective action:
\begin{equation}
S={1\over{2\kappa^2_2}}\int d^2x\sqrt{-g}e^{\delta\phi}\left[{\cal R}_g
+\gamma(\partial\phi)^2+\Lambda\right],
\label{2dimactdl}
\end{equation}
where the $\kappa^2_2$ is the two-dimensional gravitational constant and the 
cosmological constant $\Lambda$ is given by
\begin{equation}
\Lambda={{D-3}\over{2\hat{\mu}^2}}\left[2(D-2)-{{4(D-3)}\over\Delta}\right].
\label{coscnst}
\end{equation}

To bring the action (\ref{2dimactdl}) to the standard form of the 
two-dimensional dilaton gravity action, one redefines the dilaton as 
$\Phi=e^{\delta\phi}$ and then applies the Weyl rescaling of the 
metric $g_{\mu\nu}=\Phi^{-{\gamma\over{\delta^2}}}e^{\Phi\over 2}
\tilde{g}_{\mu\nu}$.  The resulting action has the following form 
\cite{youm}:
\begin{equation}
S={1\over{2\kappa^2_2}}\int d^2x\sqrt{-\tilde{g}}\left[\Phi
{\cal R}_{\tilde{g}}+{1\over 2}\partial_{\mu}\Phi\partial^{\mu}\Phi
+\Lambda\Phi^{1-{\gamma\over{\delta^2}}}e^{\Phi\over 2}\right].
\label{2dstdrdact}
\end{equation}
The kinetic term for the dilaton $\Phi$ can be removed by applying 
one more Weyl rescaling $\tilde{g}_{\mu\nu}=e^{-{\Phi\over 2}}
\bar{g}_{\mu\nu}$, resulting in the following action \cite{youm}:
\begin{equation}
S={1\over{2\kappa^2_2}}\int d^2x\sqrt{-\bar{g}}\left[\Phi{\cal R}_{\bar{g}}
+\Phi^{1-{\gamma\over{\delta^2}}}\Lambda\right].
\label{nokinact}
\end{equation}
Particularly interesting special cases of this action are the JT model 
for the $(D,a)=(4,1/\sqrt{3})$ case and the CGHS model for the $(D,a)=
(4,1)$ case.  

The field equations of the action (\ref{nokinact}) are
\begin{eqnarray}
{\cal R}_{\bar{g}}+\left(1-{\gamma\over{\delta^2}}\right)
\Phi^{-{\gamma\over{\delta^2}}}\Lambda&=&0,
\cr
\nabla_{\mu}\nabla_{\nu}\Phi-\bar{g}_{\mu\nu}{1\over 2}
\Phi^{1-{\gamma\over{\delta^2}}}\Lambda&=&0.
\label{fldeqs}
\end{eqnarray}
In the Schwarzschild gauge, the general time-dependent solution to these 
field equations takes the following form
\footnote{The most general solution in two-dimensional dilaton gravity 
theory with general dilaton potential is previously constructed in Ref. 
\cite{birk}.}:
\begin{eqnarray}
ds^2&=&-\left[{{\delta^2}\over{2\delta^2-\gamma}}\left({{x}\over{\ell}}
\right)^{2-{\gamma\over{\delta^2}}}-2\ell M\right]d\tau^2+\left[{{\delta^2}
\over{2\delta^2-\gamma}}\left({{x}\over{\ell}}\right)^{2-{\gamma\over
{\delta^2}}}-2\ell M\right]^{-1}dx^2,
\cr
\Phi&=&{x\over\ell},
\label{gensol}
\end{eqnarray}
where $\ell\equiv 1/\sqrt{\Lambda}$ and the diffeomorphism invariant 
parameter $M$ \cite{mann} defined in the following is the mass of the solution:
\begin{equation}
M=-{1\over{2\ell}}\left[(\nabla\Phi)^2\ell^2+{{\delta^2}\over{2\delta^2
-\gamma}}\Phi^{2-{\gamma\over{\delta^2}}}\right].
\label{massdef}
\end{equation}

The thermodynamic properties of the solution (\ref{gensol}) is determined 
by the behaviour of the solution at the event horizon.  The event horizon 
is located at the root of $\bar{g}_{\tau\tau}(\Phi_H)=0$, namely at
\begin{equation}
\Phi_H=\left[2\ell M\left(2-{\gamma\over{\delta^2}}\right)
\right]^{{\delta^2}\over{2\delta^2-\gamma}}.
\label{evenhor}
\end{equation}
At the event horizon, the Killing vector $k^{\mu}=\ell\eta^{\mu\nu}
\Phi_{,\nu}$ is null due to the definition (\ref{massdef}) of the mass $M$:
\begin{equation}
\left.|k|^2\right|_{\Phi_H}=\left.-\ell^2|\nabla\Phi|^2\right|_{\Phi_H}=
\left.2\ell M-{{\delta^2}\over{2\delta^2-\gamma}}\Phi^{2-{\gamma\over
{\delta^2}}}\right|_{\Phi_H}=\bar{g}_{\tau\tau}(\Phi_H)=0.
\label{killingnull}
\end{equation}
The surface gravity $\kappa$, which determines the Hawking temperature 
$T_H={\kappa\over{2\pi}}$ and is defined by $\kappa^2=\left.-{1\over 2}
\nabla^{\mu}k^{\nu}\nabla_{\mu}k_{\nu}\right|_{\Phi_H}$, is given by
\begin{equation}
\kappa={1\over{2\ell}}\Phi^{1-{\gamma\over{\delta^2}}}_H=
{1\over{2\ell}}\left[2\ell M\left(2-{\gamma\over{\delta^2}}\right)
\right]^{{\delta^2-\gamma}\over{2\delta^2-\gamma}}.
\label{surfgrav}
\end{equation}
Given the above expressions for the mass $M$ and the surface gravity 
$\kappa$ of the solution, one can see by using the first law of the 
thermodynamics that the Bekenstein-Hawking entropy is 
\begin{equation}
S_{BH}={{2\pi}\over{\kappa^2_2}}\Phi_H={{2\pi}\over{\kappa^2_2}}
\left[2\ell M\left(2-{\gamma\over{\delta^2}}\right)
\right]^{{\delta^2}\over{2\delta^2-\gamma}}.
\label{bhent2d}
\end{equation}

We now show that the Bekenstein-Hawking entropy (\ref{bhent2d}) of 
the two-dimensional black hole compactified from the dilatonic 
black hole in $D\geq 4$ is the same as the Bekenstein-Hawking entropy 
(\ref{bhentddim}) of the original dilatonic black hole in $D\geq 4$. 
One can bring the two-dimensional part $g_{\mu\nu}=G^d_{\mu\nu}$ 
($\mu,\nu=t,r$) of the near-horizon metric (\ref{neardualmet}) to the 
form (\ref{gensol}) of the solution of the 2-dimensional dilaton gravity 
theory with the action (\ref{nokinact}) by applying the Weyl rescaling 
$g_{\mu\nu}=\Phi^{-{\gamma\over{\delta^2}}}\bar{g}_{\mu\nu}=
e^{-{\gamma\over\delta}}\bar{g}_{\mu\nu}$ and then redefining the 
coordinates in the following way:
\begin{equation}
t={{|\delta|}\over\sqrt{2\delta^2-\gamma}}\tau,\ \ \ \ \ \ \ \ \ 
r=\hat{\mu}\left({x\over\ell}\right)^{{2\Delta}\over{(D-2)^2a^2}}.
\label{coordtrn}
\end{equation} 
The resulting metric $\bar{g}_{\mu\nu}$ has the form (\ref{gensol}) with the 
mass $M$ given by
\begin{equation}
M={{\delta^2}\over{2\ell(2\delta^2-\gamma)\sinh^2\alpha}}.
\label{massexprsn}
\end{equation}
By plugging the expression (\ref{massexprsn}) for the mass $M$ into the 
expression (\ref{bhent2d}) for the entropy of the two-dimensional solution 
(\ref{gensol}), making use of the following relation:
\begin{equation} 
\kappa^2_2={{\kappa^2_D}\over{\hat{\mu}^{D-2}V_{S^{D-2}}}}=
{{\kappa^2_D}\over{m^{{D-2}\over {D-3}}\sinh^{2{{D-2}\over{D-3}}}\alpha\, 
V_{S^{D-2}}}},
\label{plnkcnst}
\end{equation}
one obtains the following expression for the 
entropy:
\begin{equation}
S_{BH}={{2\pi}\over{\kappa^2_2}}\left({1\over{\sinh^2\alpha}}
\right)^{{\delta^2}\over{2\delta^2-\gamma}}=
{{2\pi}\over{\kappa^2_D}}m^{{D-2}\over{D-3}}V_{S^{D-2}}\sinh^{4\over\Delta}
\alpha.
\label{entddimfr2d}
\end{equation}
Note, the Einstein gravitational constant $\kappa_D$ is related to the 
Newton constant $G_D$ as $\kappa^2_D=8\pi G_D$ in the unit $c=1$.  So, in the 
limit of large $\alpha$ (i.e. the large charge limit or the near-extremal 
limit), in which $\sinh\delta\approx \cosh\delta$, the Bekenstein-Hawking 
entropy (\ref{entddimfr2d}) of the two-dimensional solution (\ref{gensol}) 
with (\ref{massexprsn}) becomes exactly same as the Bekenstein-Hawking 
entropy (\ref{bhentddim}) of the $D$-dimensional dilatonic black hole 
(\ref{eindil0sol}).  Therefore, thermodynamics of the $D$-dimensional 
dilatonic black hole solution (\ref{eindil0sol}) can be effectively 
described by thermodynamics of the 2-dimensional black hole solution 
(\ref{gensol}).

\section{Dilatonic $p$-Branes}

When all the longitudinal directions are compactified on a compact manifold, 
a $p$-brane in $D$ spacetime dimensions reduces to a dilatonic black hole 
in $D-p$ spacetime dimensions.  So, by using the result of the previous 
section, one can see that thermodynamics of dilatonic $p$-branes can also 
be effectively described by black holes in two-dimensional dilaton 
gravity models.  Also, thermodynamics of the delocalized intersecting 
brane solutions can be described by thermodynamics of black holes in 
two-dimensional dilaton gravity theories, since all the delocalized 
intersecting brane solutions reduce to multi-charged black holes after all the 
longitudinal and the relative transverse directions are compactified.  In the 
following, we relate a dilatonic $p$-brane in $D$ spacetime dimensions to 
a dilatonic black hole in $D-p$ spacetime dimensions. 

The Einstein-frame action for the $D$-dimensional dilatonic $p$-brane 
with an arbitrary dilaton coupling parameter $b$ is given by
\begin{equation}
S^p_E={1\over{2\kappa^2_D}}\int d^Dx\sqrt{-G^E}\left[{\cal R}_{G^E}-
{4\over{D-2}}(\partial\phi)^2-{1\over{2\cdot(p+2)!}}e^{2b\phi}F^2_{p+2}\right],
\label{einpbrnact}
\end{equation}
where $F_{p+2}$ is the field strength of the $(p+1)$-form potential 
$A^{(p+1)}=A_{M_1...M_{p+1}}dx^{M_1}\wedge...\wedge dx^{M_{p+1}}$ 
($M_1,...,M_{p+1}=0,1,...,D-1$).  The non-extreme dilatonic $p$-brane 
solution to the field equations of this action has the following form:
\begin{eqnarray}
ds^2_E&=&H^{-{{4(D-p-3)}\over{(D-2)\Delta_p}}}_p\left[-f_pdt^2+dx^2_1+\cdots+
dx^2_p\right]+H^{{{4(p+1)}\over{(D-2)\Delta_p}}}_p\left[f^{-1}_pdr^2+r^2
d\Omega^2_{D-p-2}\right],
\cr
e^{\hat{\phi}}&=&H^{{(D-2)b}\over{2\Delta_p}}_p,\ \ \ \ \ \ \ 
A_{tx_1...x_p}=-H^{-1}_p,
\label{einpbrnsol}
\end{eqnarray}
where
\begin{eqnarray}
H_p&=&1+{{m\sinh^2\alpha_p}\over{r^{D-p-3}}},\ \ \ \ 
f_p=1-{m\over{r^{D-p-3}}},
\cr
\Delta_p&=&{{(D-2)b^2}\over{2}}+{{2(p+1)(D-p-3)}\over{D-2}}.
\label{pbrndefs}
\end{eqnarray}

The Bekenstein-Hawking entropy $S_{BH}$ of the dilatonic $p$-brane solution 
(\ref{einpbrnsol}) is determined by the surface area $A_H$ of the 
horizon (located at $r=r_H=m^{1\over{D-p-3}}$):
\begin{equation}
S_{BH}={{A_H}\over{4G_D}}={{m^{{D-p-2}\over{D-p-3}}V_{S^{D-p-2}}
\cosh^{{4(p+1)(D-p-2)}\over{(D-2)\Delta_p}}\alpha_p}\over{4G_D}}.
\label{bhentpbr}
\end{equation}

Since the $p$-brane solution (\ref{einpbrnsol}) does not depend on 
the longitudinal coordinates $x_i$ ($i=1,...,p$), i.e., has the isometry 
along these directions, one can compactify the solution along the 
longitudinal directions on $T^p$ to obtain a black hole solution in 
$D-p$ spacetime dimensions.  Such dimensional reduction of the dilatonic 
$p$-brane solution (\ref{einpbrnsol}) leads to the dilatonic black hole 
solution of the form (\ref{eindil0sol}) in $D-p$ spacetime dimensions 
(i.e., $D$ in the solution (\ref{eindil0sol}) is replaced by $D-p$) with 
the dilaton coupling parameter $a$ given by
\begin{equation}
a=\sqrt{{{D-p}\over{D-p-2}}b^2+{{4(D-p-3)^2p}\over{(D-2)(D-p-2)^2}}}.
\label{dilpara}
\end{equation}
The Bekenstein-Hawking entropy of such black hole solution obtained from 
the dilatonic $p$-brane through the dimensional reduction procedure has 
the following form:
\begin{equation}
S_{BH}={{A_H}\over{4G_{D-p}}}={{m^{{D-p-2}\over{D-p-3}}V_{S^{D-p-2}}
\cosh^{4\over{\Delta}}\alpha_p}\over{4G_{D-p}}},
\label{dimredbhent}
\end{equation}
where $\Delta$ is given by Eq. (\ref{bhsoldefs}) with $D$ replaced by 
$D-p$ and $a$ given by Eq. (\ref{dilpara}).  By using the fact that the 
value of $\Delta_p$ does not change under the dimensional reduction (so, 
$\Delta_p$ in Eq. (\ref{bhentpbr}) and $\Delta$ in Eq. (\ref{dimredbhent}) 
are the same) and the following relation between the $D$-dimensional Newton 
constant $G_D$ and the $(D-p)$-dimensional Newton constant $G_{D-p}$:
\begin{equation}
G_{D-p}={{G_D}\over{V_{T^p}}}={{G_D}\over{\cosh^{{4p(D-p-3)}\over
{(D-2)\Delta_p}}\alpha_p}},
\label{dtodmpnewton}
\end{equation}
where $V_{T^p}$ is the volume of $T^p$, on which the dilatonic $p$-brane is 
compactified, one can see that the entropy (\ref{dimredbhent}) of the 
dimensionally reduced black hole in $(D-p)$-dimensions is the same as the 
entropy (\ref{bhentpbr}) of the dilatonic $p$-brane (\ref{einpbrnsol}) in 
$D$-dimensions.  In the previous section, we have shown that the 
Bekenstein-Hawking entropy of the dilatonic black hole is the same as the 
Bekenstein-Hawking entropy of the corresponding two-dimensional solution 
in the limit of large charge or the near-extremal limit.  So, the 
Bekenstein-Hawking entropy (\ref{bhentpbr}) of the dilatonic $p$-brane 
(\ref{einpbrnsol}) in $D$ dimensions has to be the same as the 
Bekenstein-Hawking entropy of a black hole solution (\ref{gensol}) of 
the associated two-dimensional dilaton gravity theory in the large charge 
limit or the near-extremal limit.

\section{Non-Extreme Black Holes with Regular BPS Limit}

In this section, we consider black holes in string theories with regular 
BPS limits.  Such black holes are four-dimensional black hole with four 
charges and five-dimensional black hole with three charges.  
The previous related works (e.g. Refs. \cite{teo,cds1,cds2}) relate such 
black holes to two-dimensional charged black hole solution of McGuigan, Nappi 
and Yost \cite{mny}, which has non-trivial dilaton field as well as $U(1)$ 
gauge field.  In such works, black hole solutions which contain a charge 
associated with the gravitational wave are considered or $U$-duality 
transformations are applied to obtain solutions with the gravitational 
wave charge.  Then, one takes the limit in which only the harmonic functions 
associated with other charges take the near-horizon limit form, while the 
harmonic function associated with the gravitational wave (and the harmonic 
function of the fundamental string with the charge assumed to be equal 
to the gravitational wave charge) does not take the near-horizon form.   
In this limit, the string-frame spacetime metric is put into the form of 
the direct product of the McGuigan, Nappi and Yost black hole and a sphere 
after the coordinate transformation.  Then, the Bekenstein-Hawking entropy 
of the two-dimensional black hole becomes exactly same as the 
Bekenstein-Hawking entropy of the $D=4,5$ black holes.

When one counts the microscopic degrees of freedom associated with the black 
hole entropy by making use of two-dimensional model or three-dimensional 
model of the BTZ black hole \cite{btz}, one considers conformal theory at 
the spacetime boundary, which is associated with the gravity theory only 
and therefore does not have anything to do with string theories 
(although it is shown \cite{btzstr,btzkal,btzak} that the BTZ black hole 
solution can be embedded as a solution of string theory).  To put it another 
way, the microscopic degrees of freedom in such picture are not associated 
with the degenerate string states, whose density is invariant under the 
$U$-duality transformations.  We also note that the form of the 
Einstein-frame spacetime metric, which gives rise to the Bekenstein-Hawking 
entropy formula, is insensitive to different ways of embedding black holes 
as the higher-dimensional intersecting branes in string theories or 
M-theory.  So, it seems to be unnatural to uplift the black hole solutions 
to ten dimensions and apply series of $U$-duality transformations to make 
the black hole solutions carry the charge associated {\it specifically} 
with the gravitational wave of string theory for the purpose of relating 
the near-horizon limit spacetime metric to the BTZ black hole solution 
(in spacetime in one higher dimensions) or to the two-dimensional charged 
black hole solution of McGuigan, Nappi and Yost.   
Also, it seems to be unnatural to let only part of harmonic functions 
take near-horizon limit forms (by applying the series of $U$-duality 
transformations and the so-called shift transformation), while that  
associated with the gravitational wave (and fundamental string) not taking 
near-horizon limit form, in order to relate the $D=4,5$ black hole solutions 
to the BTZ black hole solution (and to the two-dimensional charge black hole 
of McGuigan, Nappi and Yost), when we take notice of the fact that all the 
harmonic functions of the black hole solutions in the Einstein-frame are 
actually on the equal putting (i.e., the Einstein-frame metric is symmetric 
under the permutations of harmonic functions).   Such unnaturalness becomes 
pronounced for the particular case of the Reissner-Nordstrom black holes, 
i.e, the case of equal constituent charges.  First, when all the charges are 
equal, all the scalars of the solutions are constant, but equal charge limit 
of the above mentioned near-horizon limit solutions does not lead to constant 
scalar fields since some of harmonic functions have near-horizon form and 
some do not.  Second, it is unnatural to let some of harmonic functions not 
take near-horizon forms, when all the charges have the same magnitude.  

So, in this section, we consider only the generic Einstein-frame spacetime 
metric for black hole solutions, disregarding higher-dimensional origin of 
constituent charges and taking all the harmonic functions on the equal 
putting.  We will therefore not uplift the $D=4,5$ black holes to higher 
dimensions and we will let all the harmonic functions associated with the 
constituent charges take the near-horizon limit forms.  Perhaps, our 
description of non-extreme black hole with regular BPS limit in this section 
may not lead to the correct description of black hole thermodynamics.  But 
it seems to be more natural from the perspective of the Einstein-frame form 
of the spacetime metric of black hole solutions.  The generic property of 
stringy black holes with regular BPS limit is that in the near-horizon limit 
all the scalar fields (including dilaton) of the black hole solutions become 
constant and the spacetime metric takes the AdS$_2\times S^n$ form.   
Therefore, it seems that the solution of the associated two-dimensional 
model should have constant dilaton field. 

The most general action for the two-dimensional dilaton gravity, which 
depends at most on two derivatives of the fields, can be transformed to 
the following form \cite{banks,mgk}:
\begin{equation}
S={1\over{2\kappa^2_2}}\int d^2x\sqrt{-g}\left[\phi{\cal R}_g+V(\phi)\right].
\label{2dimdilact}
\end{equation}
The field equations of this action has a solution with 
constant dilaton $\phi=\phi_0$, provided that the potential $V(\phi)$ 
satisfies the following conditions \cite{cnst}:
\begin{equation}
V(\phi_0)=0,\ \ \ \ \ \ \ 
\left.{{dV(\phi)}\over{d\phi}}\right|_{\phi_0}\neq 0.
\label{potcond}
\end{equation}
Then, in the conformal gauge with the following spacetime metric:
\begin{equation}
g_{\mu\nu}dx^{\mu}dx^{\nu}=e^{2\rho}\left(-dt^2+dx^2\right),
\label{cnfgagmet}
\end{equation}
the following field equations of the action (\ref{2dimdilact}) for a 
static configuration:
\begin{eqnarray}
{{d^2\rho}\over{dx^2}}+{1\over 2}e^{2\rho}{{dV}\over{d\phi}}&=&0,
\cr
{{d^2\phi}\over{dx^2}}-e^{2\rho}V&=&0,
\label{fldeqn2dim}
\end{eqnarray}
lead to the solution with constant spacetime curvature, i.e. 
$\phi=\phi_0$ and ${\cal R}_g=2e^{-2\rho}{{d^2\rho}\over
{dx^2}}=-V^{\prime}(\phi_0)$.  This is in accordance with the fact that the 
near-horizon region spacetimes of the four-dimensional black hole with four 
charges and the five-dimensional black hole with three charges contain the 
AdS$_2$ space.  The following spacetime metric solution \cite{cnst} to the 
field equations (\ref{fldeqn2dim}) in the Schwarzschild gauge is obtained 
by redefining the spatial coordinate through the relation $dy=e^{2\rho}dx$:
\begin{equation}
ds^2_2=-\left({{{\cal R}_0}\over 2}y^2-k\right)dt^2+
\left({{{\cal R}_0}\over 2}y^2-k\right)^{-1}dy^2,
\label{cnstdilmet}
\end{equation}
where ${\cal R}_0=-V^{\prime}(\phi_0)$ is the Ricci scalar of the metric 
and $k$ is an integration constant.

In the following subsections, we bring the near-horizon region metrics of 
the $D=4,5$ black holes to the form of the solution (\ref{cnstdilmet}).  
Then, the study of thermodynamics of non-extreme black holes in $D=4,5$ 
with regular BPS limit reduces to the study of two-dimensional black holes 
with constant dilaton field.

\subsection{Four-dimensional black hole}

The generic form of the Einstein-frame metric of the four-dimensional 
black hole solution with four charges is
\begin{equation}
g^E_{\mu\nu}dx^{\mu}dx^{\nu}=-{1\over\sqrt{H_1H_2H_3H_4}}fdt^2+
\sqrt{H_1H_2H_3H_4}\left[f^{-1}dr^2+r^2d\Omega^2_2\right],
\label{4dimeinmet}
\end{equation}
where $f=1-{m\over r}$ and $H_i=1+{{m\sinh^2\alpha_i}\over r}$.
Here, the $U(1)$ charges $Q_i\sim m\sinh 2\alpha_i$ can have any 
higher-dimensional origin.  Namely, the metric (\ref{4dimeinmet}) can be 
the metric of a heterotic black hole solution \cite{hetbh1,hetbh2,hetbh3} 
with the Kaluza-Klein $U(1)$ electric and magnetic charges and the NS-NS 
2-form $U(1)$ electric and magnetic charges or the metric of a type-IIB 
black hole solution compactified from intersecting D3-branes.  Regardless of 
various higher-dimensional origins as intersecting branes, the Einstein-frame 
metric for all the four-dimensional stringy black hole with four charges 
have the form (\ref{4dimeinmet}).  

In the near-horizon region, in which $H_i\approx{{m\sinh^2\alpha_i}\over
{r}}$ ($i=1,...,4$), the metric (\ref{4dimeinmet}) is approximated to
\begin{eqnarray}
g^{E}_{\mu\nu}dx^{\mu}dx^{\nu}&\approx&-{{r^2}\over{m^2\prod^4_{i=1}
\sinh\alpha_i}}\left(1-{m\over r}\right)dt^2+
{{m^2\prod^4_{i=1}\sinh\alpha_i}\over{r^2}}\left(1-{m\over r}
\right)^{-1}dr^2
\cr
& &+m^2\prod^4_{i=1}\sinh\alpha_id\Omega^2_2.
\label{nearhor4dimein}
\end{eqnarray}
The two-dimensional part of this near-horizon region metric can be put into 
the following suggestive form of the constant dilaton solution 
(\ref{cnstdilmet}) of the two-dimensional dilaton gravity by redefining the 
spatial coordinate as $y=r-{m\over 2}$:
\begin{eqnarray}
ds^2_2&\approx&-\left({{y^2}\over{m^2\prod^4_{i=1}\sinh\alpha_i}}
-{1\over{4\prod^4_{i=1}\sinh\alpha_i}}\right)dt^2
\cr
& &+\left({{y^2}\over{m^2\prod^4_{i=1}\sinh\alpha_i}}
-{1\over {4\prod^4_{i=1}\sinh\alpha_i}}\right)^{-1}dy^2.
\label{2dimof4dim}
\end{eqnarray}
This corresponds to two-dimensional solution with constant dilaton and 
constant spacetime curvature ${\cal R}_g=-V^{\prime}(\phi_0)=2/
(m^2\prod^4_{i=1}\sinh\alpha_i)$.

\subsection{Five-dimensional black hole}

The generic form of the Einstein-frame metric of the five-dimensional 
black hole solution in string theory with three charges, regardless of the 
higher-dimensional origins of charges, is as follows:
\begin{equation}
g^E_{\mu\nu}dx^{\mu}dx^{\nu}=-{1\over{\left(H_1H_2H_3\right)^{2\over 3}}}fdt^2
+\left(H_1H_2H_3\right)^{1\over 3}\left[f^{-1}dr^2+r^2d\Omega^2_3\right],
\label{5dimeinmet}
\end{equation}
where $f=1-{m\over{r^2}}$ and $H_i=1+{{m\sinh^2\alpha_i}\over{r^2}}$.
In the near-horizon region, in which $H_i\approx{{m\sinh^2\alpha_i}\over
{r^2}}$ ($i=1,2,3$), the metric (\ref{5dimeinmet}) is approximated to
\begin{eqnarray}
g^{E}_{\mu\nu}dx^{\mu}dx^{\nu}&\approx&-{{r^4}\over{m^2\prod^3_{i=1}
\sinh^{4/3}\alpha_i}}\left(1-{m\over{r^2}}\right)dt^2+{{m\prod^3_{i=1}
\sinh^{2/3}\alpha_i}\over{r^2}}\left(1-{m\over{r^2}}\right)^{-1}dr^2
\cr
& &+m\prod^3_{i=1}\sinh^{2/3}\alpha_id\Omega^2_3.
\label{nearhor5dimein}
\end{eqnarray}

The two-dimensional part of the metric (\ref{nearhor5dimein}) can be put 
into the following suggestive form of the two-dimensional constant dilaton 
solution (\ref{cnstdilmet}) by redefining the spatial coordinate 
as $y={1\over{2m^{1/2}\prod^3_{i=1}\sinh^{1/3}\alpha_i}}(r^2-{m\over 2})$:
\begin{eqnarray}
ds^2_2&\approx& -\left({{4y^2}\over{m\prod^3_{i=1}\sinh^{2/3}\alpha_i}}
-{1\over{4\prod^3_{i=1}\sinh^{4/3}\alpha_i}}\right)dt^2
\cr
& &+\left({{4y^2}\over{m\prod^3_{i=1}\sinh^{2/3}\alpha_i}}
-{1\over{4\prod^3_{i=1}\sinh^{4/3}\alpha_i}}\right)^{-1}dy^2.
\label{2dimof5dim}
\end{eqnarray}
This corresponds to the two-dimensional solution with the constant dilaton 
and the constant spacetime curvature ${\cal R}_g=-V^{\prime}(\phi_0)
=8/(m\prod^3_{i=1}\sinh^{2\over 3}\alpha_i)$.


\begin{thebibliography} {99}
\small
\parskip=0pt plus 2pt

\bibitem{pol} J. Polchinski, ``Dirichlet-branes and Ramond-Ramond charges,''
Phys. Rev. Lett. {\bf 75} (1995) 4724, hep-th/9510017.

\bibitem{sv} A. Strominger and C. Vafa, ``Microscopic origin of the 
Bekenstein-Hawking entropy,'' Phys. Lett. {\bf B379} (1996) 99, hep-th/9601029.

\bibitem{tho} G. 't Hooft, ``Dimensional reduction in quantum gravity,''
gr-qc/9310026.

\bibitem{sus1} L. Susskind, ``Strings, black holes and Lorentz contraction,''
Phys. Rev. {\bf D49} (1994) 6606, hep-th/9308139.

\bibitem{sus2} L. Susskind, ``The World as a hologram,'' J. Math. Phys. 
{\bf 36} (1995) 6377, hep-th/9409089.

\bibitem{bh} J.D. Brown and M. Henneaux, ``Central charges in the canonical 
realization of asymptotic symmetries: an example from three-dimensional 
gravity,'' Commun. Math. Phys. {\bf 104} (1986) 207.

\bibitem{car1} S. Carlip, ``The statistical mechanics of the 
(2+1)-dimensional black hole,'' Phys. Rev. {\bf D51} (1995) 632, 
gr-qc/9409052.

\bibitem{car2} S. Carlip, ``The statistical mechanics of the 
three-dimensional Euclidean black  hole,'' Phys. Rev. {\bf D55} (1997) 
878, gr-qc/9606043.

\bibitem{str} A. Strominger, ``Black hole entropy from near-horizon 
microstates,'' JHEP {\bf 02} (1998) 009, hep-th/9712251.

\bibitem{btz} M. Banados, C. Teitelboim and J. Zanelli, ``The black hole in 
three-dimensional space-time,'' Phys. Rev. Lett. {\bf 69} (1992) 1849, 
hep-th/9204099.

\bibitem{hyu} S. Hyun, ``$U$-duality between three and higher dimensional 
black holes,'' hep-th/9704005.

\bibitem{ss} K. Sfetsos and K. Skenderis, ``Microscopic derivation of the 
Bekenstein-Hawking entropy formula for non-extremal black holes,'' Nucl. 
Phys. {\bf B517} (1998) 179, hep-th/9711138.

\bibitem{teo} E. Teo, ``Statistical entropy of charged two-dimensional black 
holes,'' Phys. Lett. {\bf B430} (1998) 57, hep-th/9803064.

\bibitem{cds1} G.L. Cardoso, ``Charged heterotic black-holes in four and two 
dimensions,'' Phys. Lett. {\bf B432} (1998) 65, hep-th/9804064.

\bibitem{cds2} G.L. Cardoso and T. Mohaupt, ``Dual heterotic black holes in 
four and two dimensions,'' Phys. Lett. {\bf B435} (1998) 277, hep-th/9806036.

\bibitem{klp} Y. Kiem, C. Lee and D. Park,
``Thermodynamics of doubly charged CGHS model and D1 - D5 - KK black 
holes of IIB supergravity,'' Phys. Rev. {\bf D58} (1998) 125002, 
hep-th/9806182.

\bibitem{cdn} M. Cadoni, ``Dimensional reduction of 4D heterotic string 
black holes,'' Phys. Rev. {\bf D60} (1999) 084016, hep-th/9904011.

\bibitem{jac} R. Jackiw, ``Liouville field theory: a two-dimensional model 
for gravity?,'' MIT-CTP-1049, in {\it Quantum theory of gravity}, ed. S. 
Christensen (Adam Hilgar, Bristol, 1984) p. 403.

\bibitem{tei} C. Teitelboim, ``The Hamiltonian structure of two-dimensional 
space-time and its relation with the conformal anomaly,'' Print-83-0130 
(Texas), in {\it Quantum theory of gravity}, ed. S. Christensen (Adam Hilgar, 
Bristol, 1984) p. 327.

\bibitem{cghs} C.G. Callan, S.B. Giddings, J.A. Harvey and A. Strominger, 
``Evanescent black holes,'' Phys. Rev. {\bf D45} (1992) 1005, hep-th/9111056.

\bibitem{mny} M.D. McGuigan, C.R. Nappi and S.A. Yost, ``Charged black holes 
in two-dimensional string theory,'' Nucl. Phys. {\bf B375} (1992) 421, 
hep-th/9111038.

\bibitem{cdm1} M. Cadoni and S. Mignemi, ``Entropy of 2D black holes from 
counting microstates,'' Phys. Rev. {\bf D59} (1999) 081501, hep-th/9810251.

\bibitem{cdm2} M. Cadoni and S. Mignemi, ``Asymptotic symmetries of AdS$_2$  
and conformal group in d=1,'' hep-th/9902040.

\bibitem{nav} J. Navarro-Salas and P. Navarro, ``AdS$_2$/CFT$_1$ 
correspondence and near-extremal black hole entropy,'' hep-th/9910076.

\bibitem{card} J.L. Cardy, ``Operator content of two-dimensional conformally 
invariant theories,'' Nucl. Phys. {\bf B270} (1986) 186.

\bibitem{carl} S. Carlip, ``Black hole entropy from conformal field theory 
in any dimension,'' Phys. Rev. Lett. {\bf 82} (1999) 2828, hep-th/9812013.

\bibitem{sol} S.N. Solodukhin, ``Conformal description of horizon's states,'' 
Phys. Lett. {\bf B454} (1999) 213, hep-th/9812056.

\bibitem{dl} M.J. Duff and J.X. Lu, ``Black and super $p$-branes in 
diverse dimensions,'' Nucl. Phys. {\bf B416} (1994) 301, hep-th/9306052.

\bibitem{gt} G.W. Gibbons and P.K. Townsend, ``Vacuum interpolation in 
supergravity via super $p$-branes,'' Phys. Rev. Lett. {\bf 71} (1993) 3754, 
hep-th/9307049.

\bibitem{dgt} M.J. Duff, G.W. Gibbons and P.K. Townsend, ``Macroscopic 
superstrings as interpolating solitons,'' Phys. Lett. {\bf B332} (1994) 321, 
hep-th/9405124.

\bibitem{bst} H.J. Boonstra, K. Skenderis and P.K. Townsend, ``The domain 
wall/QFT correspondence,'' JHEP {\bf 01} (1999) 003, hep-th/9807137.

\bibitem{youm} D. Youm, ``(Generalized) conformal quantum mechanics of 
0-branes and two-dimensional dilaton gravity,'' hep-th/9909180.

\bibitem{birk} D. Louis-Martinez and G. Kunstatter, ``On Birckhoff's theorem 
in two-dimensional dilaton gravity,'' Phys. Rev. {\bf D49} (1994) 5227.

\bibitem{mann} R.B. Mann, ``Conservation laws and 2-D black holes in 
dilaton gravity,'' Phys. Rev. {\bf D47}, 4438 (1993), hep-th/9206044.

\bibitem{btzstr} G.T. Horowitz and D.L. Welch, ``Exact three-dimensional 
black holes in string theory,'' Phys. Rev. Lett. {\bf 71} (1993) 328, 
hep-th/9302126.

\bibitem{btzkal} N. Kaloper, ``Miens of the three-dimensional black hole,''
Phys. Rev. {\bf D48} (1993) 2598, hep-th/9303007.

\bibitem{btzak} A. Ali and A. Kumar, ``$O(\tilde{d},\tilde{d})$ 
transformations and 3D black hole,'' Mod. Phys. Lett. {\bf A8} (1993) 
2045, hep-th/9303032.

\bibitem{banks} T. Banks and M. O'Loughlin, ``Two-dimensional quantum gravity 
in Minkowski space,'' Nucl. Phys. {\bf B362} (1991) 649.

\bibitem{mgk} D. Louis-Martinez, J. Gegenberg and G. Kunstatter, 
``Exact Dirac quantization of all 2-D dilaton gravity theories,'' 
Phys. Lett. {\bf B321} (1994) 193, gr-qc/9309018.

\bibitem{cnst} J. Cruz, A. Fabbri, D.J. Navarro and J. Navarro-Salas, 
``Integrable models and degenerate horizons in two-dimensional gravity,'' 
hep-th/9906187.

\bibitem{hetbh1} M. Cveti\v c and D. Youm, ``Dyonic BPS saturated black holes 
of heterotic string on a six torus,'' Phys. Rev. {\bf D53} (1996) 584, 
hep-th/9507090.

\bibitem{hetbh2} M. Cveti\v c and D. Youm, ``BPS saturated and non-extreme 
states in Abelian Kaluza-Klein theory and effective N = 4 supersymmetric 
string vacua,'' hep-th/9508058.

\bibitem{hetbh3} M. Cveti\v c and D. Youm, ``All the static spherically 
symmetric black holes of heterotic string on a six torus,'' Nucl. Phys. 
{\bf B472} (1996) 249, hep-th/9512127.


\end{thebibliography}
\end{document}